\documentclass[intlimits,twoside,a4paper]{article}

\usepackage[cp1251]{inputenc}

\usepackage{bbding}
\usepackage[eqsecnum]{cmpj3}

\usepackage{bm}

\newcommand{\bea}{\begin{eqnarray}}
\newcommand{\eea}{\end{eqnarray}}
\newcommand{\be}{\begin{equation}}
\newcommand{\ee}{\end{equation}}
\newcommand{\eps}{\varepsilon}

\issue{2022}{25}{4}{43710}
\doinumber{10.5488/CMP.25.43710}

\title[Electric field induced polarization rotation in squaric acid crystals revisited]
{Electric field induced polarization rotation in squaric acid crystals revisited}

\author{A. P. Moina\orcid{0000-0002-2170-2445}}

\address{Institute for Condensed Matter Physics of the National Academy of Sciences of Ukraine,\\
	1 Svientsitskii St., 79011 Lviv, Ukraine}


\date{Received July 10, 2022, in final form July 26, 2022}
\begin{document}

\maketitle

\begin{abstract}
Using the previously developed model we revisit the problem of the electric field induced polarization rotation in antiferroelectric crystals of squaric acid. We test an alternative set of the model parameters, according to which the dipole moments associated with the H$_2$C$_4$O$_4$ groups are assumed to be parallel to the diagonals of the $ac$ plane. The $T$-$E$ phase diagrams and the polarization curves $P(E)$ for the fields directed along the $a$ axis and along one of the diagonals are considered. Comparison of the theoretical results with the newly published experimental data confirm the validity of the model. The calculations reveal no apparent advantage of the new set of the parameters over the previously used set.
\printkeywords
\end{abstract}


\maketitle


\section{Introduction}

The squaric acid H$_2$C$_4$O$_4$ is a classical two-dimensional antiferroelectric. The crystal is tetragonal,~$I4/m$, in the paraelectric phase and monoclinic,  $P2_1/m$, in the antiferroelectric phase. The hydrogen bonded C$_4$O$_4$ groups form sheets, parallel to the $ac$ plane and stacked along the $b$-axis. Below the transition  at 373~K, a spontaneous polarization arises in these sheets, with the neighboring sheets polarized in the opposite directions \cite{semmingsen:95,semmingsen:77,hollander:77}.

External electric fields applied to a uniaxial antiferroelectric  can switch a sublattice polarization by 180$^\circ$ and induce thereby the transition from  antiferroelectric (AFE) to ferroelectric (FE) phase. The (pseudo)tetragonal symmetry of the squaric acid crystal lattice and of its hydrogen bond networks allows the sublattice polarizations to be directed along two perpendicular axes in the fully ordered system. As a result, here the external field can rotate one of the sublattice polarizations by 90$^\circ$, whereupon a noncollinear ferrielectric phase with perpendicular sublattice polarizations (NC90 \cite{moina:21}) is induced. The possibility of such a rotation has been suggested by Horiuchi et al \cite{horiuchi:18}, and their hysteresis loop measurements and Berry phase calculations   gave evidence for it. Further calculations \cite{ishibashi:18} indicated that the 90$^\circ$ rotation is possible at different orientations of the field within the $ac$ plane. It is also predicted~\cite{horiuchi:18,ishibashi:18} that application of higher fields along the diagonals of the $ac$ plane can lead to the second rotation of the negative sublattice polarization by 90$^\circ$ and induction of the collinear ferroelectric phase.

Recently \cite{moina:21,moina:21:2} we  developed a deformable \cite{moina:20} two-sublattice proton-ordering model  for a description of squaric acid behaviour in external electric fields, applied arbitrarily within the plane of  hydrogen bonds.
The model calculations confirm the two-step process of polarization reorientation~\cite{horiuchi:18,ishibashi:18} at low temperatures, with the negative sublattice polarization being switched twice by 90$^\circ$ at each transition,
for any orientation of the field within the $ac$ plane, but a few exceptional directions. The exceptional directions are those, when the field is either i) collinear to the axes of the sublattice polarization in the AFE phase, or ii) directed at $45^\circ$ to these axes. In the case
i), the crystal behaves like a uniaxial antiferroelectric, undergoing a single-step polarization switching to the FE phase without the intermediate noncollinear phase, while in the case ii), the transition field from the NC90 to the FE phase goes to infinity, i.e., the transition never occurs \cite{moina:21:2}. 

The temperature-electric field  phase diagrams of squaric acid were constructed \cite{moina:21,moina:21:2} for the field $E_1$($E_3$) directed along the $a$($c$) tetragonal axis, for the fields denoted for brevity as $E_1\pm E_3$ and directed along the diagonals of the $ac$ plane, as well as for the fields of the two  above mentioned exceptional directions i) and ii).  Note that the $T$-$E$ diagrams are identical for the fields rotated by $90^\circ$ around the $b$ axis, because of the pseudotetragonal symmetry of the model \cite{moina:21:2}.

Experimentally, the low-temperature transition between the NC90 and FE phases has not been detected yet due to the dielectric breakdown of the samples. As follows from the model calculations \cite{moina:21:2}, the field of this transition is the lowest when its direction is close to the axis of the sublattice polarization, so it is most likely to be experimentally observed at this field orientation. 

On the other hand, for the AFE-NC90 switching, the experimental data by Horiuchi {et al} \cite{horiuchi:18} had been available, when our calculations were carried out. The polarization hysteresis curves at different temperatures for the field $E_1$  had been measured, and the temperature dependence of the switching field had been deduced from those; for the field $E_1+E_3$, the measurements had been performed for one temperature only \cite{horiuchi:18}. With the fitting procedure for the model being based on the data \cite{horiuchi:18} for the static dielectric permittivity, the obtained agreement between the theory and the  experiment for the switching fields and for the $P(E)$ curves was only qualitative \cite{moina:21,moina:21:2}. Quantitatively, the agreement was conspicuously unsatisfactory, which led us to believe that the model used  was not completely appropriate, and that essential modifications were required \cite{moina:21:2}. Quite recently, however, the same group of Horiuchi {et al} reported \cite{horiuchi:21} the results of their new  measurements of the polarization loops for the squaric acid crystals of an improved dielectric strength. This permitted to increase the maximum electric field that could be applied to the samples in the hysteresis experiments. Our preliminary calculations showed that the new experimental data were much closer to the predictions of the model \cite{moina:21,moina:21:2} than the previous data of  \cite{horiuchi:18}, and that the doubts concerning the model validity were premature. 

It was extensively discussed in \cite{moina:21} that the accepted set of the values of the model parameters, in particular of the dipole moments assigned to the 
ground state configurations of the H$_2$C$_4$O$_4$ groups, is not unique.  While the magnitude of the dipole moment vector is constrained by the fitting to the  permittivity~\cite{horiuchi:18}, its orientation (and thereby the orientation of the ground state sublattice polarizations) can be varied within the $ac$ plane. With the set of the model parameters adopted in \cite{moina:21,moina:21:2}  these vectors are oriented at about $56^\circ$ to the $a(c)$ axes. On the other hand, the Berry phase calculations \cite{horiuchi:18,horiuchi:21} indicate that the axes of the spontaneous sublattice polarization, in fact, are very close or even coincide with the diagonals of the  $ac$ plane. In terms of our model, this  means that the crystallographic axes and the diagonals are the  above mentioned exceptional directions: the axis $a (c)$  is the direction ii), while the diagonals of the $ac$ plane  are the direction i).
The topology of the $T$-$E$ diagrams and the shape of the $P(E)$ curves for the fields $E_1 (E_3)$ and $E_1\pm E_3$ will change accordingly. The availability of the new, more reliable experimental data \cite{horiuchi:21} makes a quantitative comparison of  theoretical and experimental $P(E)$ curves meaningful and could help to ascertain the orientation of the model dipole moment vectors. 

Thus, it seems  worthwhile to revisit the problem of polarization rotation in squaric acid,  to perform calculations with an alternative set of the model parameters, where the sublattice polarizations are oriented along the diagonals of the $ac$ plane, and to compare the theoretical results with the most recent~\cite{horiuchi:21} experimental data. The model \cite{moina:21,moina:21:2}, briefly described in section~\ref{themodel}, is used without any further modification of the formulae. In section~\ref{calculations} the results of the theoretical calculations with the old and new sets of the model parameters are compared with the experimental data.

\section{The model}
\label{themodel}

The model has been introduced and explicated in \cite{moina:21}, and a concise outline is given in \cite{moina:21:2}. Below we present a brief qualitative description of the model; all the formulae and other
relevant details and discussions can be found in the mentioned papers.

\begin{figure}[!t]
	\centerline{
				\includegraphics[width=\columnwidth]{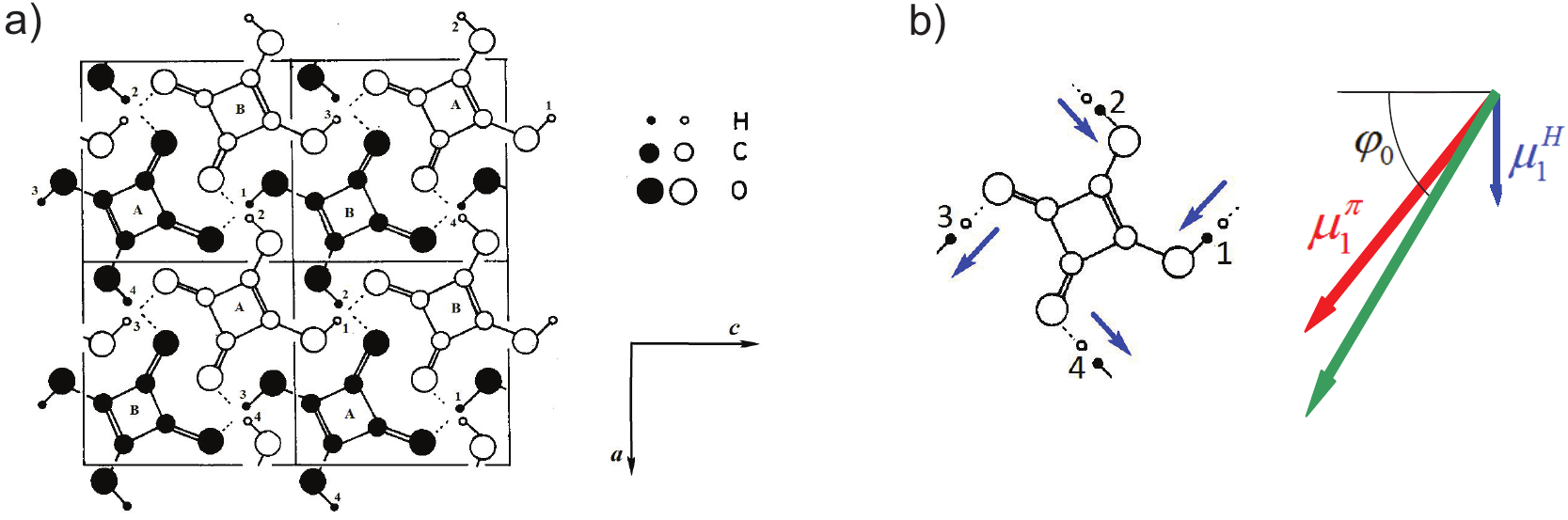}
}
	\caption{(Colour online) a) The crystal structure of squaric acid as viewed along the $b$ axis.  Two adjacent layers are shown, with black and open circles each. The A and B type C$_4$O$_4$ groups are indicated (see \cite{moina:20,moina:21} for explanation), and the hydrogen bonds are numbered, $f=1,2,3,4$. b) The dipole moments assigned to one of the four lateral proton configurations (the configuration 1 in tables~1 in \cite{moina:21,moina:21:2}).  Directions of the dipole moments associated with protons ${\bm{\mu}}^H_1=(2\mu^H,0,0)$ and with electrons $\bm{\mu}^\pi_1=(2\mu^\pi_\parallel,0,-2\mu^\pi_\perp)$ are shown with blue and red arrows, respectively; the green arrow is the total dipole moment of the configuration; the vector lengths are nominal. $\varphi_0=\arctan (\mu^H+\mu_{\parallel}^\pi)/\mu_{\perp}^\pi$ is the angle between the total dipole moment of configuration 1 and the $c$ axis. Figures are taken from \cite{semmingsen:74,moina:20,moina:21,moina:21:2}.} \label{outline}
\end{figure}

Protons on the hydrogen bonds in squaric acid move in double-well potentials, so each of the protons can occupy one of  the two sites on the bond: closer to the given C$_4$O$_4$ group or to the neighboring group. The motion of protons is described by Ising pseudospins, whose two eigenvalues are assigned to two equilibrium positions of each proton. Two interpenetrating sublattices (layers) of pseudospins are considered.

The total system Hamiltonian \cite{moina:21,moina:21:2} includes ferroelectric intralayer long-range interactions between pseudospins, ensuring ferroelectric ordering within each separate layer, antiferroelectric interlayer interactions responsible for alternation of polarizations in the stacked layers, and the short-range interactions, which include also the coupling with external electric fields $E_1$ and $E_3$ directed along the tetragonal (paraelectric) $a$ and $c$ axes of the crystal.

The short-range Hamiltonian describes the four-particle configurational correlations between protons placed around each C$_4$O$_4$ group. The usual Slater-Takagi type scheme \cite{matsushita:80,matsushita:82,moina:20,moina:21} of 16 degenerate levels of lateral/diagonal/single-ionized/double-ionized proton  configurations is assumed. The lateral and  single-ionized configurations have dipole moments in the $ac$ plane; the degeneracy of their energy levels is removed by the electric fields $E_1$ and $E_3$, which break the equivalence of the hydrogen bonds that link the C$_4$O$_4$ groups along the $a$ and $c$ axes (see tables~1 in \cite{moina:21,moina:21:2}). 

Assignment of the dipole moments to the ground-state lateral configurations is the crucial point of the model. We rely on the results of the Berry phase calculations \cite{horiuchi:18}, which have shown that the ground-state sublattice polarization in this crystal is formed directly by displacements of protons along the hydrogen bonds and, mostly, by the electronic contributions of switchable $\pi$-bond dipoles.  

Positions of the $\pi$-bonds are determined by the proton arrangement around the given C$_4$O$_4$ group: in the lateral configurations the $\pi$-bond is formed between the two neighboring carbons, near which protons sit on the hydrogen bonds (see fig.~\ref{outline}b), and also between the carbons and adjacent to them oxygens, next to which there is no proton (meaning that the protons on these H-bonds sit in the minima close to the neighboring C$_4$O$_4$ groups).
%
The field-induced polarization rotation by $90^\circ$ or $180^\circ$ occurs via  flipping of one or two protons in each molecule to the other sites along the same hydrogen bonds and via a simultaneous switching of the $\pi$-bonds. For the depicted in figure~\ref{outline}b lateral proton configuration, the vector of the proton contribution to the dipole moment is oriented along the $a$ axis, while the electronic contribution is at the angle to this axis. The dipole moments of the three remaining lateral configurations can be obtained from the scheme of figure~\ref{outline}b by rotation by a multiple of 90$^\circ$.

After going from the representation of proton configuration energies to the pseudospin representation, the four-particle cluster approximation for the obtained short-range Hamiltonian is employed. The mean field approximation is used for the long-range interlayer and intralayer interactions \cite{moina:20,moina:21}.
 The dependence of all proton-proton interaction parameters on
the diagonal components of the lattice strain tensor and on the H-site distance, which are changed by the thermal expansion and potentially by an external stress if such is applied, is taken into account \cite{moina:20}. The expression for the thermodynamic potential has been obtained \cite{moina:21}; the order parameters and lattice strains are found by numerical minimization thereof.




The values of all model parameters were chosen earlier \cite{moina:20,moina:21,moina:21:2}. In particular, they were required~\cite{moina:20} to provide the best fit to the experimental temperature curves of the order parameter at ambient pressure, to the temperature and hydrostatic pressure dependences of the diagonal lattice strains, and to the pressure dependence of  the transition temperature $T_{\textrm N}$ in squaric acid. 

The dielectric characteristics and other electric field effects  in our model are mostly governed by values of the dipole moments, which
enter the final expressions only via the sum $\mu^H+\mu_\parallel^\pi$ and via $\mu_\perp^\pi$. These values are found by fitting the calculated curve of the static dielectric permittivity $\eps_{11}$  at zero external bias field to the experimental points of \cite{horiuchi:18}, while  trying to get the best possible agreement with the experiment for the values of the switching fields, corresponding to the first 90$^\circ$ rotation of the sublattice polarization by the field $E_1$. It can be shown that in the paraelectric phase $\eps_{11}\sim \bar\mu^2$, where
\[\bar\mu=\sqrt{(\mu^H+\mu_\parallel^\pi)^2+(\mu_\perp^\pi)^2}\]
is half the magnitude of the dipole moment, assigned to the H$_2$C$_4$O$_4$ groups.
It means that above $T_{{\textrm N}}$ the permittivity $\eps_{11}$ at zero field is determined by the magnitude of the dipole moment vector only, whereas the orientation of the  vector within the $ac$ plane can be varied.  For the set, adopted in \cite{moina:21,moina:21:2} and presented in table~\ref{tbl1} as the set A, the dipole moment and the ground state sublattice polarization are oriented at the angle $\varphi_0=\arctan (\mu^H+\mu_{\parallel}^\pi)/\mu_{\perp}^\pi\approx56^\circ$ to the crystallographic axes. However, the results of the Berry phase calculations \cite{horiuchi:21} indicate that the angle should be closer to $45^\circ$. Thus, we find an alternative set of the dipole moment values with $\mu^H+\mu_{\parallel}^\pi=\mu_{\perp}^\pi$ and with the same $\bar\mu$ as in the set A, which yields the same fit to the permittivity in the paraelectric phase; this is the set B in table~\ref{tbl1}. In the next section, using the set B, we construct the $T$-$E$ phase diagrams and explore the $P(E)$ curves for the electric fields $E_1$ and $E_1+E_3$. The results are compared with the previous calculations \cite{moina:21:2} performed with the set A, as well with the experimental data of \cite{horiuchi:18,horiuchi:21}.

\begin{table}[htb]
	\caption{The adopted values of the model dipole moments. The set A is  taken from \cite{moina:21:2}. The values of all other model parameters are the same as in \cite{moina:20,moina:21,moina:21:2}. }
	\label{tbl1}
	\vspace{0.5cm}
	\begin{center}
		\renewcommand{\arraystretch}{0}
		
		\begin{tabular}{cccc}
			\hline
& $\mu^H+\mu^\pi_\parallel$ & $\mu^\pi_\perp$ & $\bar \mu$ \\
			& \multicolumn{3}{c}{($10^{-29}$~C m)} \\
			\hline
			set A & 3.16 &2.12 & 3.8\strut 	\\ 	
			set B  &2.66 &2.66 & 3.8 \strut 	\\ 	
			 \hline
		\end{tabular}

		\vspace{1ex}
		\renewcommand{\arraystretch}{1}
	\end{center}
\end{table}


\section{Calculations}
\label{calculations}

\subsection{Phase diagrams}
\begin{figure}[hbt]
	\centerline{\includegraphics[height=0.4\columnwidth]{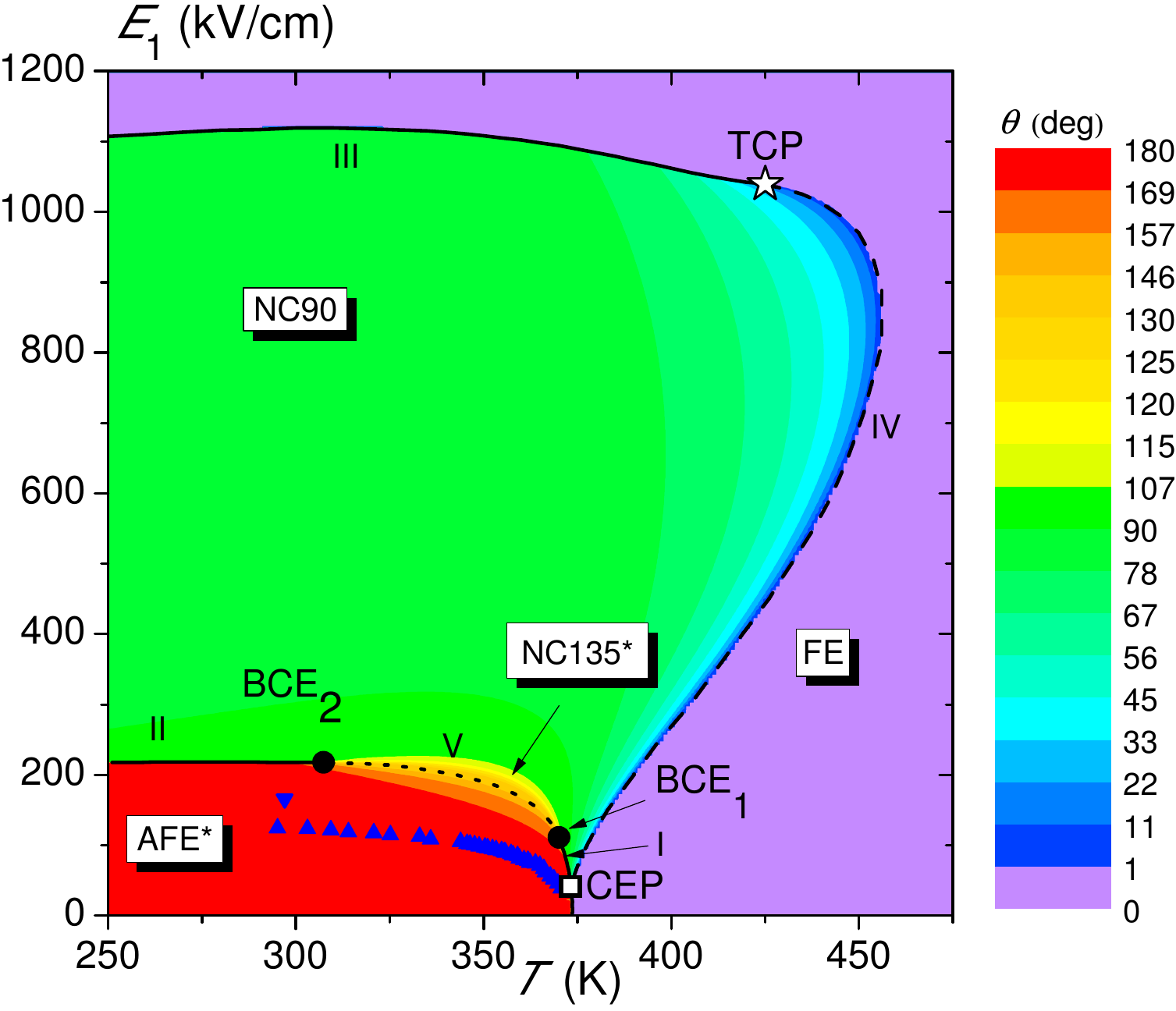}~~~~\includegraphics[height=0.4\columnwidth]{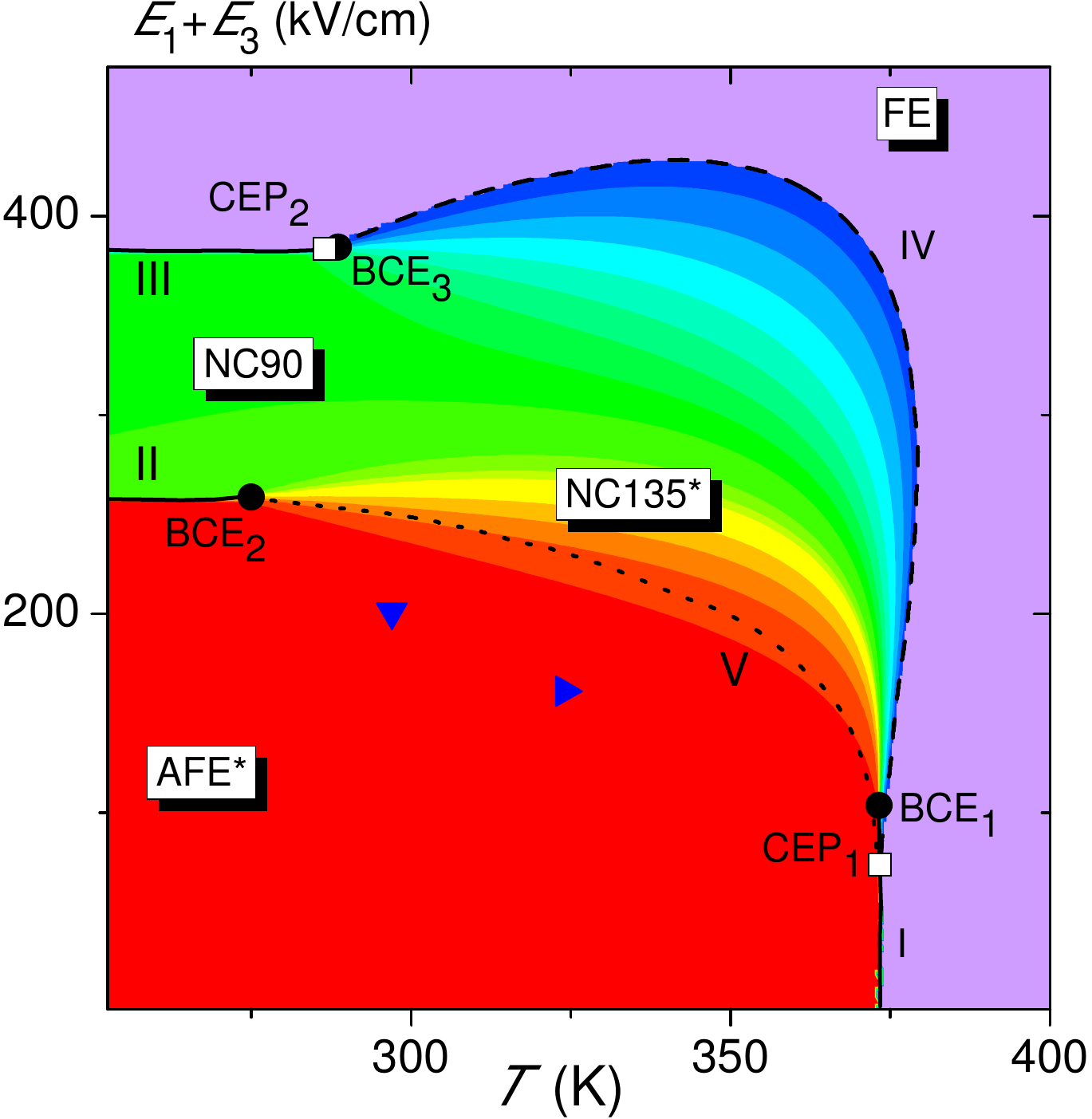}}
 	\caption{(Colour online) The $T$-$E$ phase diagrams of the squaric acid, overlapping the $T$-$E$ color contour plots of the noncollinearity angle $\theta$. The set A is used in calculations. Solid and dashed lines indicate the first and second order phase transitions, respectively; dotted lines are the supercritical lines, corresponding to the loci of maxima in the field dependences of $\textrm{d}P(E)/\textrm{d}E$. The open squares {\large $\square$}, star \FiveStarOpen, and full circles {\large $\bullet$} indicate the critical end points (CEP), tricritical point (TCP), and bicritical end points (BCE), respectively. Blue full triangles $\blacktriangle$, $\blacktriangleright$, and $\blacktriangledown$ are the experimental points of \cite{horiuchi:18}, the electronic supplementary material thereto, and \cite{horiuchi:21}, respectively.} \label{pdthetaE}
 \end{figure}

In figure~\ref{pdthetaE} we redraw the $T$-$E$ diagrams of squaric acid for   the fields $E_1$ and $E_1+ E_3$,  obtained earlier in \cite{moina:21,moina:21:2} along with the newly available experimental points of \cite{horiuchi:21}. Here, the set A of the dipole moment values was used in the calculations. The diagrams overlap the color gradient plots of the introduced in~\cite{moina:21} noncollinearity angle
$\theta$, which is the angle between the vectors of the sublattice polarizations.

Different phases in the diagrams are separated by the lines of the first order phase transitions I, II, and III, and of the second order phase transitions IV. All these lines terminate at various critical points (bicritical end points BCE, tricritical point TCP, critical end points CEP). Some of the critical points can be artifacts of the mean field approximation, used for the long-range interactions. This was discussed extensively in \cite{moina:21,moina:21:2}; we shall not dwell on this here.  
The phase denoted as AFE* (the red region) is non-collinear antiferrielectric, very close to the initial AFE phase with $\theta\sim 180^\circ$. The purple region is the collinear field-induced ferroelectric phase (FE) with $\theta=0$. The phase between the transition lines II, III, and IV (green and blue) is the noncollinear ferrielectric phase NC90, where $\theta$ mostly remains close to~90$^\circ$, only rapidly decreasing to zero near the second-order phase transition line IV. In the region NC135* (orange to yellow), a crossover between the AFE* and NC90 phases occurs. Here,  $\theta$ changes gradually from $\sim180^\circ$ to $\sim90^\circ$: the negative sublattice polarization rotates continuously with increasing field and becomes perpendicular to the positive sublattice polarization. As discussed in \cite{moina:21,moina:21:2}, this continuous rotation is a statistically averaged effect, possible only in presence of thermal fluctuations. 

Crossovers are often marked by the lines formed by the loci of the extrema of the response functions~--- second derivatives of the thermodynamic potentials. Those supercritical lines are continuations of the first order transition lines beyond the critical points terminating them. The major drawback of this method is that the extrema of different response functions yield different supercritical lines; moreover, the supercritical lines formed by the extrema of the same response function taken along different thermodynamic paths (e.g., isotherms or isofields) are different as well (see \cite{schienbein:18}). In order to compare the theory and the experimental data derived from the field dependence of polarization, we  mark the crossover between the AFE* and NC90 phases using the lines formed by the maxima of the $\textrm{d}P(E)/\textrm{d}E$ isotherms (the inflection points of the $P(E)$ isotherms), where $P$ is the projection of the net polarization vector on the field axis. These are the dotted lines V in the phase diagrams. 


As one can see in the left-hand panel of figure~\ref{pdthetaE}, for the field $E_1$, the most recent data obtained in~\cite{horiuchi:21} for the sample with the improved dielectric strength appear to be in a much better agreement with the theory  than the earlier experimental data of \cite{horiuchi:18}. The theoretical switching fields, calculated with the set A, are much higher than the experimental values of \cite{horiuchi:18} with the relative error $\eta=1-E_{\textrm{exp}}/E_{\textrm{theor}}\approx 0.42$ at room temperature (295~K). The error decreases down to $\approx0.23$ for the experimental points of  \cite{horiuchi:21}, which is still not quite satisfactory, but evidently much better. 

In the case of  $E_1 + E_3$, the switching fields calculated with the set A  are higher than predicted for the field $E_1$. This is in a qualitative agreement with all available experimental observations \cite{horiuchi:18,horiuchi:21}. The relative errors are about 0.3 for \cite{horiuchi:18} at 324~K and 0.22  for \cite{horiuchi:21} at 295~K, that is, the improvement here is not so striking.

Now let us see how the situation changes, when the set B is used in calculations. For this set, the ground state spontaneous polarization axis is oriented along the diagonal of the $ac$ plane. It means that the fields $E_1+E_3$ and $E_1$ are directed along this axis and at $45^\circ$ to it, respectively, that is, along the exceptional directions i) and ii), discussed in Introduction.  It is then expected that the $T$-$E$ diagrams will be topologically different from those, depicted in figure~\ref{pdthetaE}. For the field $E_1+E_3$, the crystal of squaric acid should behave like a uniaxial antiferroelectric, exhibiting
a one-step polarization rotation by $180^\circ$ without the intermediate noncollinear phase. For the field $E_1$, the field of switching to the ferroelectric phase is expected to tend to infinity, and only the AFE*-NC90 transition can be observed.

 \begin{figure}[hbt]
	\centerline{\includegraphics[height=0.4\columnwidth]{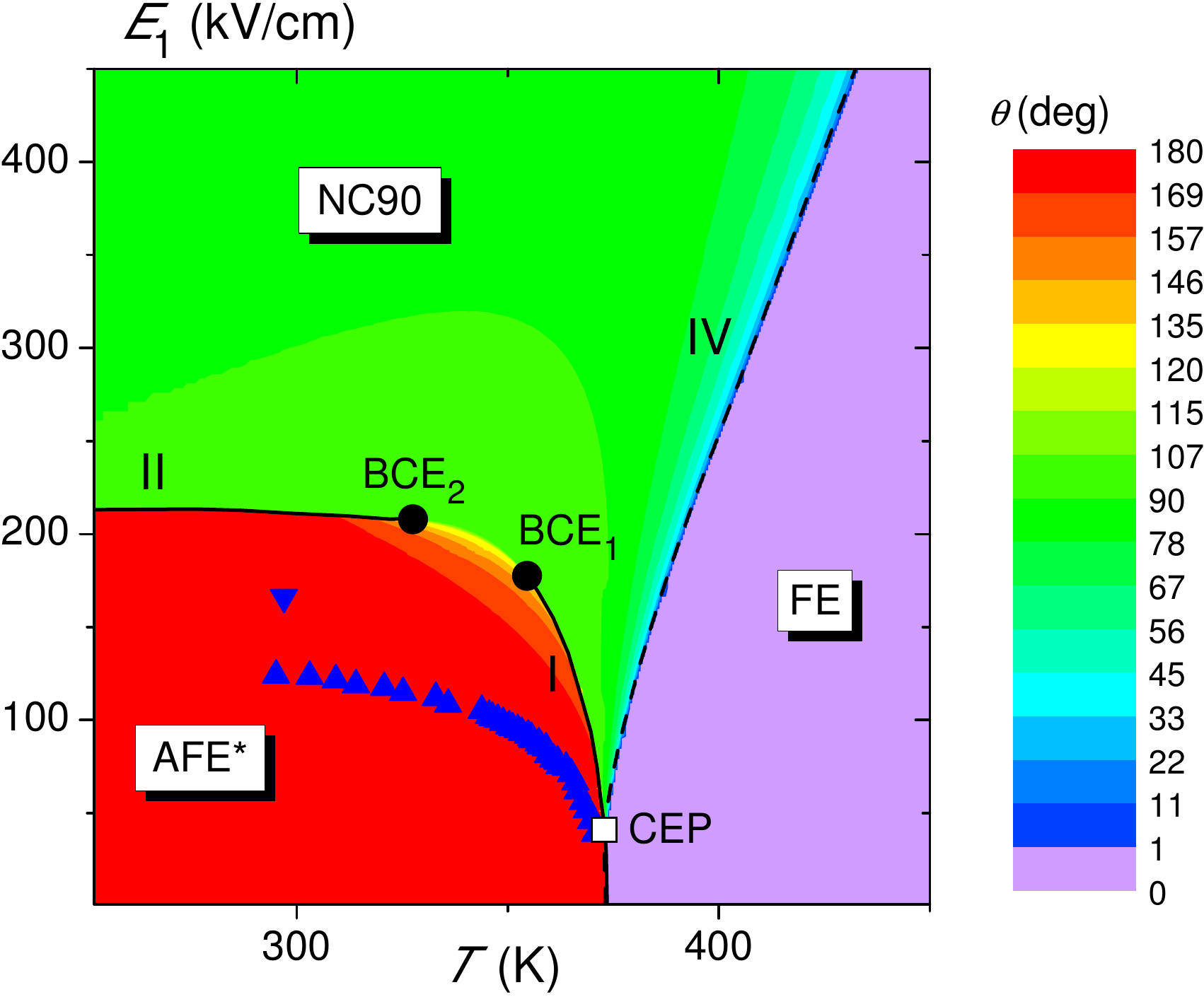}~~~~\includegraphics[height=0.4\columnwidth]{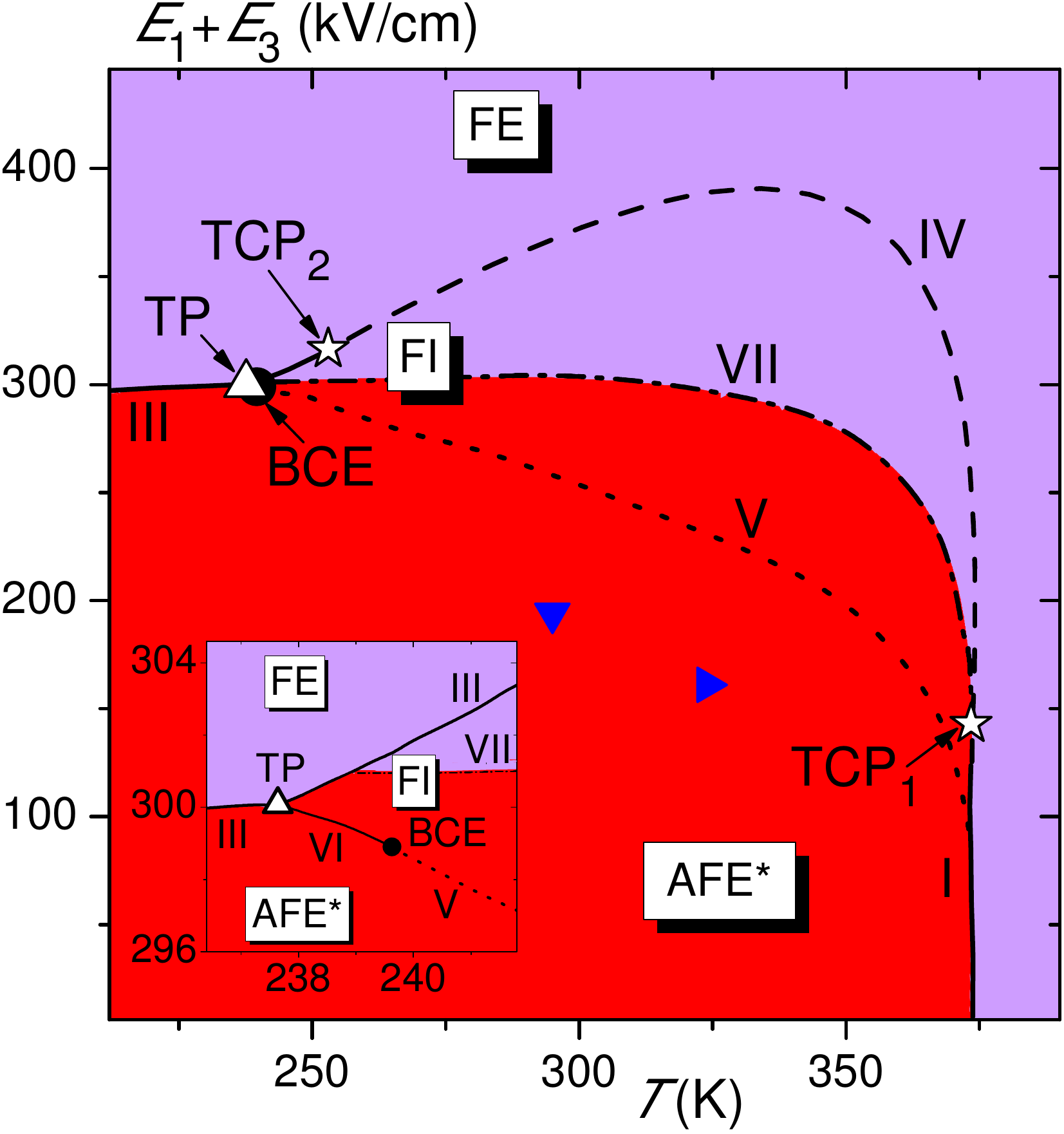}}
	\caption{(Colour online) The same as in figure~\ref{pdthetaE}. The set B is used in calculations. The open triangle $\triangle$ indicates the triple point TP. The dash-dotted line VII corresponds to the loci of minima in the field dependences of $\textrm{d}P(E)/\textrm{d}E$. The other notations are the same as in figure~\ref{pdthetaE}. } \label{pdthetaEnew}
\end{figure}

The  $T$-$E$ phase diagrams, calculated with the set B and presented in figure~\ref{pdthetaEnew}, are indeed in a total agreement with the above described  picture.  
As the magnitude of the dipole moment 2$\bar\mu$ is the same for both sets, these diagrams are numerically identical to those obtained in \cite{moina:21:2} with the set A for the exceptional directions ii) and i), respectively (see figures~8, 9 in \cite{moina:21:2}). This identity can be proved algebraically, using the expression for the thermodynamic potential of the system \cite{moina:21,moina:21:2}.

For the field $E_1$, the positions of the lines II of the AFE*-NC90 phase transitions, calculated with the sets A and B, are very close but not the same (c.f. the left-hand panels in figures~\ref{pdthetaE} and \ref{pdthetaEnew}). The closeness can be explained by the found in \cite{moina:21:2} dependence of this  switching field at low temperatures on the orientation of the spontaneous sublattice polarization axis $E^{II}\sim 1/\cos(\delta \varphi-\piup/4)$, where $\delta\varphi$ is the angle between this axis and the external field $E$. Since $\delta\varphi$ for the sets A and B differ by about $11^\circ$ only, the difference between the corresponding switching fields is small as well. It is then trivial to say that for $E_1$, the sets A and B yield about the same agreement with the experimental data for the switching field.

For the field $E_1+E_3$, the intermediate phase NC90 is absent, and
$\theta$ is always either $180^\circ$ or zero (see the right-hand panel of figure~\ref{pdthetaEnew}), i.e., all phases are collinear. The polarization switching occurs either as a first order phase transition across lines III directly to the FE phase and across line VI to an intermediate collinear ferrielectric phase FI, or gradually. In the latter case,  the magnitude of one of the sublattice polarizations decreases down to zero with increasing field, changes its sign continuously at line VII, and then increases until the second order transition to the FE phase occurs at line IV. Interestingly, line VII, where the angle $\theta$ changes from $180^\circ$ to 0, is formed by the loci of the \textit{minima} of the $\textrm{d}P(E)/\textrm{d}E$ isotherms, as opposed to line V, formed by the loci of the maxima.  Line VI, emanating from the critical point BCE (see the inset in figure~\ref{pdthetaEnew}), marks the crossover between AFE* and FI phases. It is to be compared with the experimental data for the switching fields, and it yields nearly the same agreement as the set A, with the relative errors about 0.3 for \cite{horiuchi:18} at 324~K and 0.23 for \cite{horiuchi:21}.

\subsection{Polarization}

\begin{figure}[!t]
\centerline{
\includegraphics[height=0.33\textwidth]{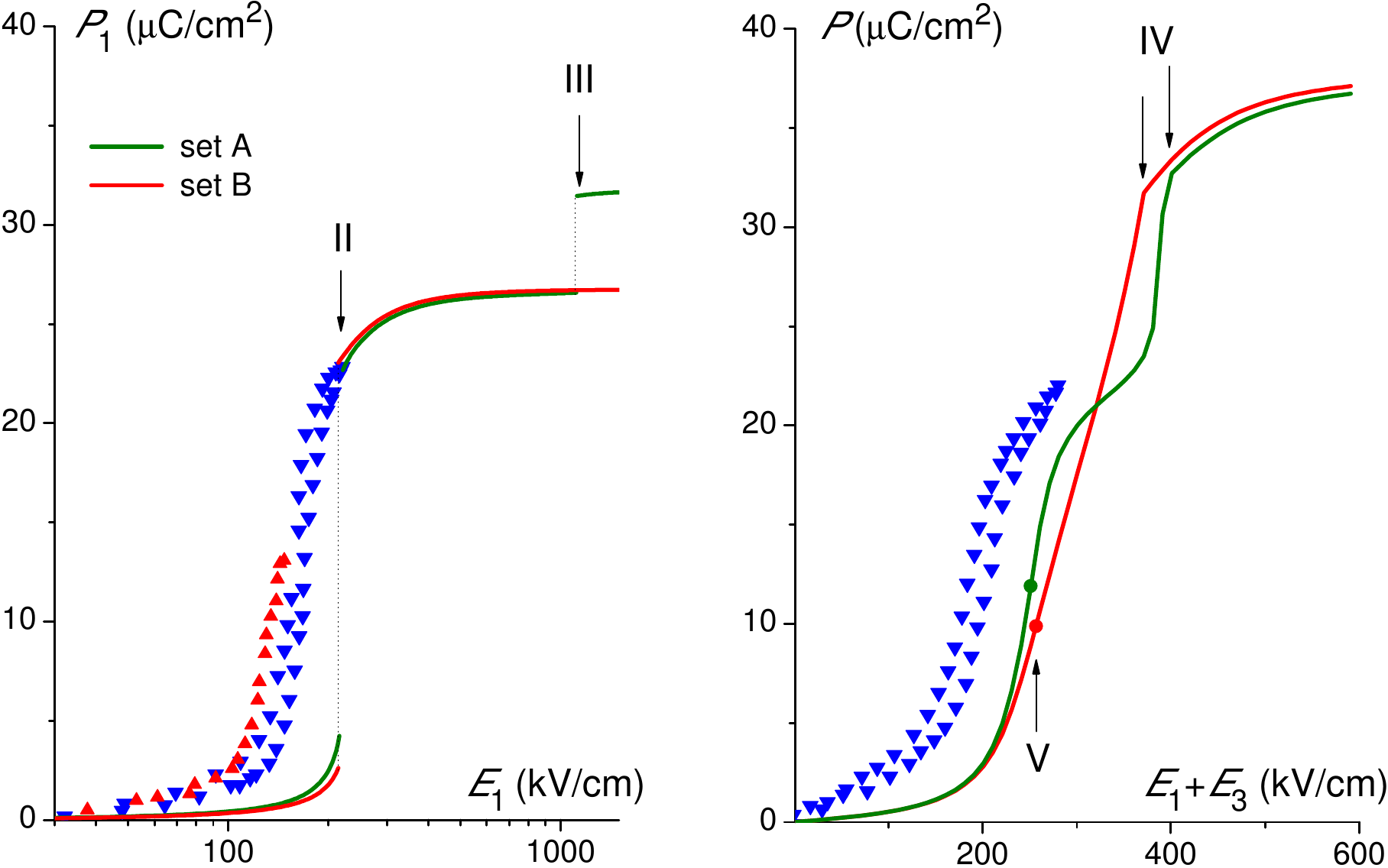}  
}		
\caption{(Colour online) The field dependences of polarizations at 295~K. Full triangles: experimental points taken from \cite{horiuchi:18} ($\blacktriangle$) and \cite{horiuchi:21} ($\blacktriangledown$). The arrows indicate phase transitions of the first order  across lines II, III (left-hand) and of the second order across lines IV (right-hand). The arrow and full circles ($\large\bullet$) indicate the crossovers at lines V (right-hand). Lines II-V are from the $T$-$E$ phase diagrams, figures~\ref{pdthetaE}, \ref{pdthetaEnew}.}  \label{pol_rr}
\end{figure}

In figure~\ref{pol_rr} we plot the field dependences of the projections of the net polarization vector on the field direction for the fields  $E_1$ and $E_1+E_3$. The experimental points of \cite{horiuchi:18} and \cite{horiuchi:21} are also presented. The drastic changes in the experimental hysteresis curves, brought by the improvement of the sample quality and by the increase of the maximum value of the applied field  in \cite{horiuchi:21}, are very well seen. It is obvious that the comparison of the theoretical polarization curves with the earlier data of \cite{horiuchi:18} could be only qualitative.

As one can see, for $E_1$, the sets A and B predict three and two plateaus of polarization, respectively. In the physically reasonable field range, which includes the AFE*-NC90 first order phase transition (at lines II from the phase diagrams \ref{pdthetaE}, \ref{pdthetaEnew}), the two sets yield very similar  polarizations. The calculated polarization jumps are 17.9~\textmu C/cm$^2$ for the set A and 20.4~\textmu C/cm$^2$ for the set B, in a fair agreement with the experimental 17.2~\textmu C/cm$^2$ \cite{horiuchi:21}.
The set A also predicts the second step of polarization at a much higher field, at the transition to the FE phase (across line III from the phase diagram, figure~\ref{pdthetaE}). It seems unlikely, however, that the field of such high a magnitude could ever be applied in an experiment without the squaric acid samples suffering the dielectric breakdown.

For the field $E_1+E_3$, the two sets of the model parameters yield different behaviour of polarization even at experimentally accessible fields.
The $P(E)$ curve, calculated with the set A, has three smeared plateaus, with a clear rounded step, corresponding to the AFE*- NC90 crossover across line V, and then a cusp at the NC90-FE second order transition across line IV. The lower part of this curve, albeit being shifted to higher fields, is in a good qualitative and quantitative agreement with the experimental points. The polarization, calculated with the set B, on the other hand, has only two smeared plateaus: at low fields and above the cusp at line IV. No pronounced intermediate plateau is seen. The change of concavity at the inflection point, marked by a full circle in the figure, is hardly discernible. The agreement with the experiment is visibly worse than for the set A. However, the switching field magnitude for $E_1+E_3$ is predicted \cite{moina:21:2,horiuchi:18,horiuchi:21} to be higher than for the field $E_1$. It means that, despite the increased dielectric strength of the samples, the maximum applied fields $E_1+E_3$ \cite{horiuchi:21} could still be insufficient to obtain correct data for polarization. A potential further improvement of the crystal quality (if such is still possible) may change the measured values of polarization and switching field for the diagonally directed field in the same way, as such an improvement did in the case of the field $E_1$ in \cite{horiuchi:21} as compared to \cite{horiuchi:18}, which is well illustrated in the left-hand panel of figure~\ref{pol_rr}. Then, the agreement with the theoretical curves can be reexamined.

\section{Concluding remarks}
Using the previously developed \cite{moina:21} deformable two-sublattice proton ordering model, we revisit the problem of polarization rotation in antiferroelectric crystals of squaric acid under the influence of external electric fields. The unique structure of the two-dimensional hydrogen bond networks in squaric acid permits 90$^\circ$ rotation of the sublattice polarization. The model predicts \cite{moina:21,moina:21:2} that except for some particular directions of the field,
the polarization reorientation at low temperatures is a two-step 
process: first, to the noncollinear phase with perpendicular sublattice polarizations and then to the collinear ferroelectric phase. However, when the field is directed along the 
axis of spontaneous sublattice polarizations, the intermediate noncollinear phase is absent; when the field is at 45$^\circ$ to this axis, the field of the transition to the ferroelectric phase tends to infinity.

The previously obtained $T$-$E$ phase diagrams and newly calculated polarization curves are compared with the most recent experimental data \cite{horiuchi:21}, measured using the crystal samples of the increased dielectric strength. We also test an alternative
set of the model parameters, for which the dipole moments assigned to the H$_2$C$_4$O$_4$ groups are of the same magnitude as in the previous calculations, but oriented along the diagonals of the $ac$ plane.

The new experimental data \cite{horiuchi:21} are in a drastically better agreement with the theory than the earlier results \cite{horiuchi:18}, especially for the polarization curves, as well as for the switching fields. It shows that the simplicity of the model was not the major reason of  the earlier \cite{moina:21,moina:21:2} disagreement  between theory and experiment and gives a strong evidence for the model validity.


Results of testing the new set of the model parameters are inconclusive. Overall, the comparison of the theoretical polarization curves with the experimental data seems to slightly favor the previous set \cite{moina:21,moina:21:2}, according to which the axes of the spontaneous sublattice polarization are close, but not exactly parallel 
to the diagonals of the $ac$ plane.
Further experimental studies may shed some light on this problem.

As far as a further verification of the model is concerned, the appropriateness of the mean field approximation, used for the long-range interactions, may be addressed. This approximation is, most likely, the origin of the artifact splitting \cite{moina:21,moina:21:2} of some tricritical points in the $P$-$E$ phase diagrams into the systems of bicritical and critical endpoints and also of the appearance of the intermediate FI phase, seen in the right-hand panels of figures~\ref{pdthetaE}, \ref{pdthetaEnew}. Monte Carlo calculations may be used to construct more accurate diagrams.


\newpage
\ukrainianpart

\title{Ще раз про обертання поляризації електричним полем  в  кристалах квадратної кислоти}
\author{А. П. Моїна}
\address{
	Інститут фізики конденсованих систем Національної академії наук України\\
	79011, м. Львів, вул. Свєнціцького, 1
}
%
%
%

\makeukrtitle

\begin{abstract}
	\tolerance=3000%
	З використанням запропонованої раніше моделі розглядаються процеси обертання поляризації зовнішніми електричними полями в антисегнетоелектричних кристалах квадратної кислоти.
    Обчислення також проводяться з альтернативним набором параметрів теорії, в якому дипольні моменти, які приписуються групам H$_2$C$_4$O$_4$, паралельні до діагоналей площини $ac$. Досліджено фазові діаграми $T$-$E$ та криві поляризації  $P(E)$ для полів, прикладених уздовж осі  $a$ та уздовж діагоналі площини  $ac$. Порівняння теоретичних результатів з нещодавно опублікованими експериментальними даними підтверджує правильність запропонованої моделі. Не виявлено суттєвої переваги нового набору параметрів моделі перед тим, що використовувався в попередніх розрахунках.
	
	\keywords
	поляризація, електричне поле, фазовий перехід, антисегнетоелектрик, фазова діаграма, квадратна кислота
	
\end{abstract}

\end{document}